\documentclass[vci-paper]{vciart}

\usepackage{graphicx}
\usepackage{amsmath}   

\usepackage{amssymb}
\usepackage[latin1]{inputenc}

\def\dj{d\hskip-.4em\hbox{\char'26}\hskip-.1em}
\def\Dj{D\hskip-.8em\lower.4ex\hbox{\char'26}\hskip.3em}
\begin{document}

\begin{frontmatter}

\title{Resolution and Efficiency of the ATLAS Muon Drift-Tube Chambers at High Background
Rates}

\author[A]{M.~Deile},
\author[B]{J.~Dubbert},
\author[D]{S.~Horvat\thanksref{E}}
\author[D]{O.~Kortner},
\author[D]{H.~Kroha},
\author[D]{A.~Manz},
\author[D]{S.~Mohrdieck-M\"ock},
\author[B]{F.~Rauscher},
\author[D]{R.~Richter},
\author[B]{A.~Staude},
\author[D]{W.~Stiller}

\address[A]{CERN, CH-1211 Geneva 23, Switzerland}
\address[B]{Ludwig-Maximilians-Universit\"at M\"unchen, Am Coulombwall 1,
	D-85748 Garching, Germany}
\address[D]{Max-Planck-Institut f\"ur Physik,
	F\"ohringer Ring 6, D-80805 M\"unchen, Germany}

\thanks[E]{On leave of absence from Institut Ru\dj er Bo\v skovi\' c, 10 001 Zagreb, Croatia.}

\begin{abstract}
The resolution and efficiency of a precision drift-tube chamber for the ATLAS 
muon spectrometer with final read-out electronics was tested at the Gamma 
Irradiation Facility at CERN in a 100~GeV muon beam and at photon irradiation rates
of up to 990~Hz/cm$^2$ which corresponds to twice the highest background rate expected 
in ATLAS. A silicon strip detector telescope was used as external reference in the beam.
The pulse-height measurement of the read-out electronics was used to
perform time-slewing corrections which lead to an improvement of the average 
drift-tube resolution from 104~$\mu$m to 82~$\mu$m without irradiation and from
128~$\mu$m to 108~$\mu$m at the maximum expected rate.
The measured drift-tube efficiency agrees with the expectation from the dead time
of the read-out electronics up to the maximum expected rate.
\end{abstract}

\end{frontmatter}


\section{Introduction}
The muon spectrometer of the ATLAS detector at the Large Hadron Collider (LHC)
is designed for the measurement of muon momenta with an accuracy of $3\%$ over a
wide energy range reaching $10\%$ resolution at 1~TeV.
The muon trajectories in the 0.4~T field of a superconducting air-core toroid magnet 
are measured in three stations of precision
drift chambers chambers, the monitored drift-tube (MDT) chambers.

The MDT chambers consist of two triple or quadruple layers
of aluminium drift tubes of 0.4~mm wall thickness and 29.170~mm
inner diameter with a $50~\mu$m diameter gold-plated tungsten-rhenium wire
at the center. The tubes are filled with Ar:CO$_2$~(93:7) gas mixture at a
pressure of 3~bar.
Operated at a gas gain of 2$\cdot$10$^4$, the drift tubes must
provide a spatial resolution of better than 100~$\mu$m in order to reach the
required chamber position resolution of better than 40~$\mu$m~\cite{TDR}
with a sense wire positioning accuracy of 20~$\mu$m which has been achieved
in the chamber serial production~\cite{MPI1}.

\begin{figure}[t]
\includegraphics[width=\linewidth]{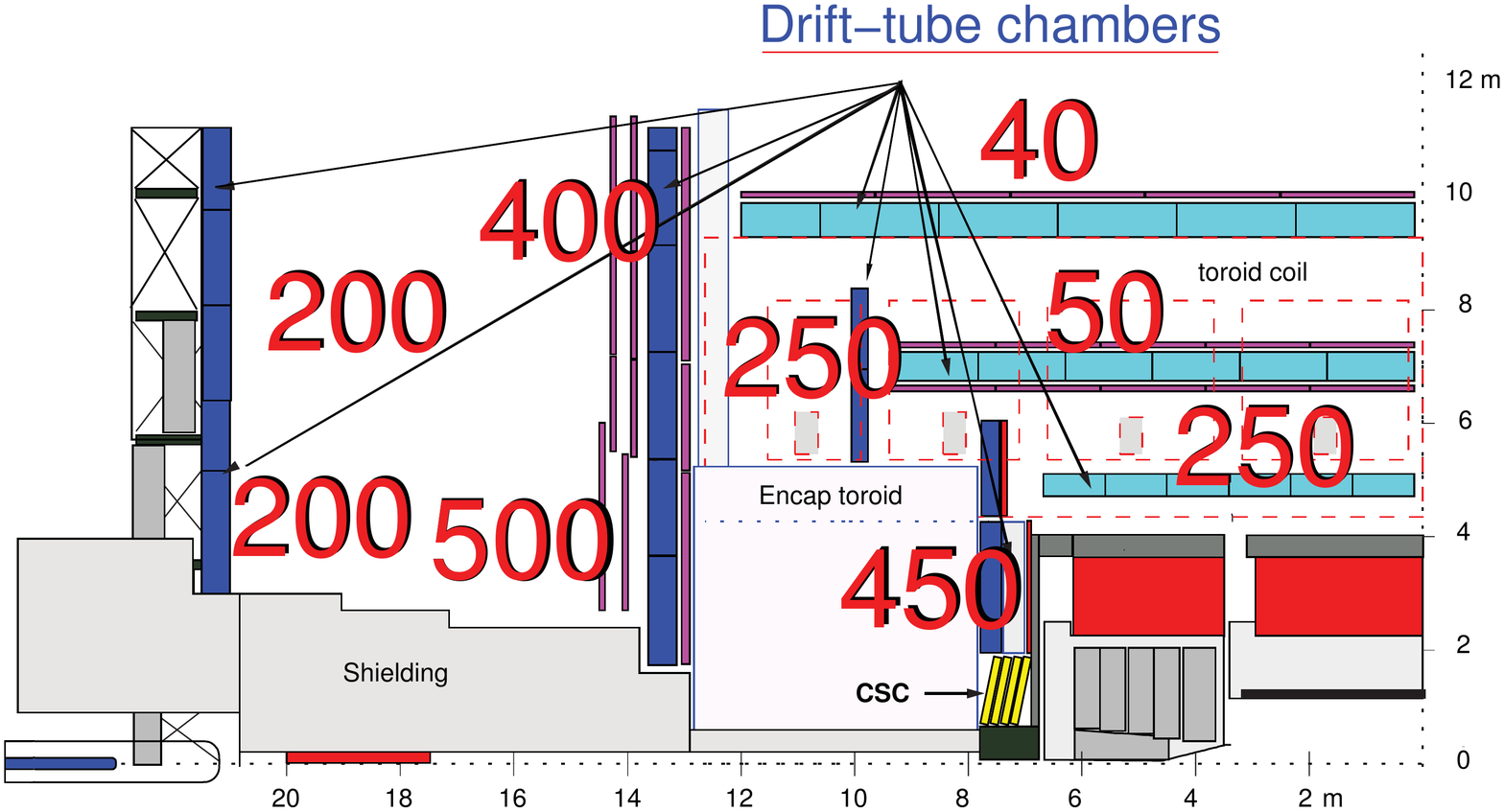}
\caption{\label{fig1}Expected background count rates in Hz/cm$^2$ (including a safety factor of 5)
in the MDT chambers in different regions of the ATLAS muon spectrometer. The cross section of
one quadrant of the ATLAS detector containing the LHC proton beam line is shown.}
\end{figure}

The operating conditions of the ATLAS muon chambers at the LHC are characterized by
unprecedentedly high neutron and $\gamma$ backgrounds (see Figure~\ref{fig1}).
The chambers will experience background count rates ranging from 40~Hz/cm$^2$
to 500~Hz/cm$^2$ including a safety factor of 5~\cite{TDR} corresponding
to count rates per tube between 45 and 300~kHz for tube lengths of $1.7-5.8$~m.

\begin{figure}[t]
\includegraphics[width=\linewidth]{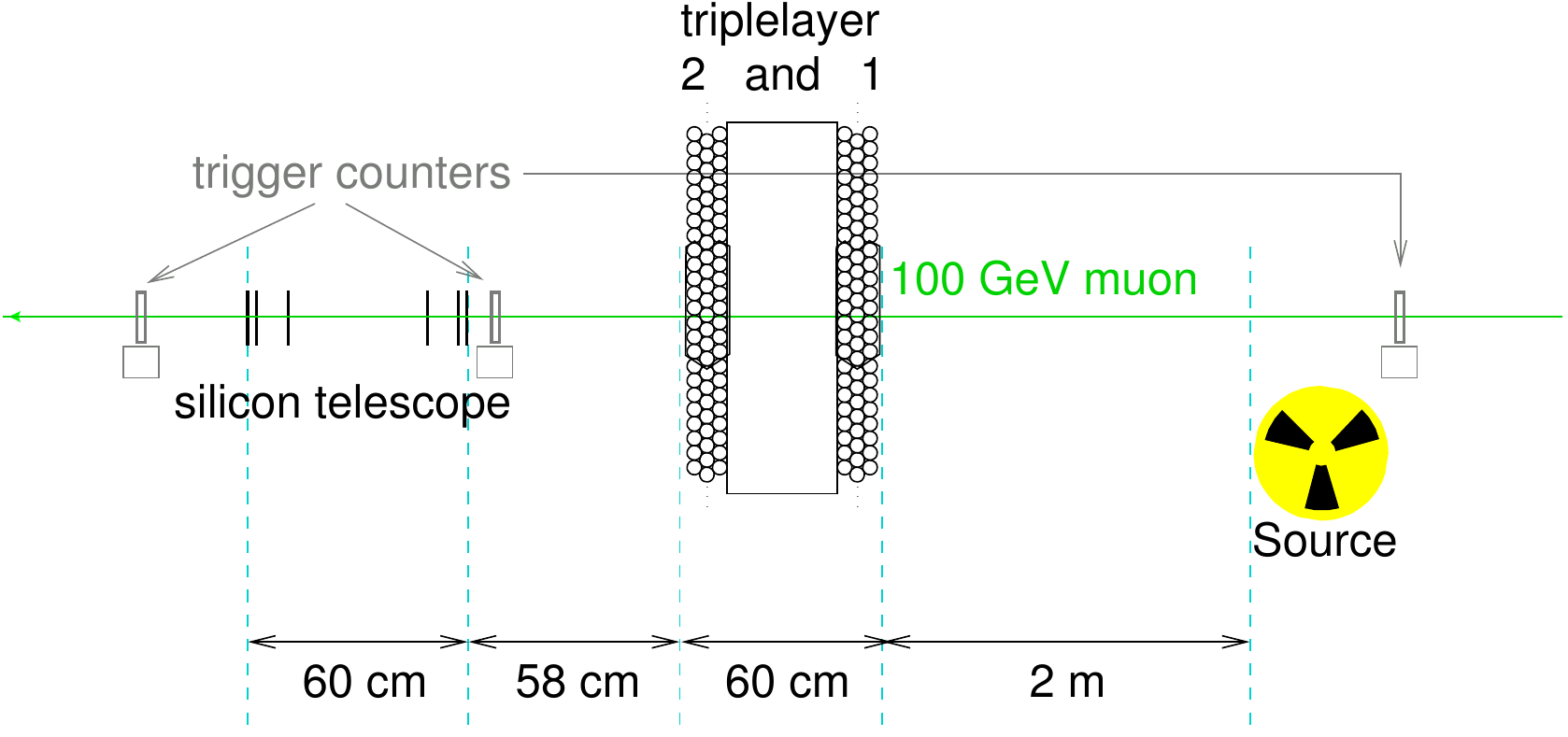}
\caption{\label{fig2}Top view of the experimental set-up at the Gamma Irradiation Facility
at CERN with a MDT chamber consisting of two triple layers of drift tubes and a
silicon strip detector telescope in a 100~GeV muon beam.}
\end{figure}

In summer 2003, one of the largest MDT chambers constructed for the ATLAS muon
spectrometer containing 432 drift tubes of 3.8~m length~\cite{MPI1}
has been tested at the Gamma Irradiation Facility~\cite{GIF} at CERN with a
740~GBq $^{137}$Cs source in a 100~GeV muon beam. The chamber was equipped with
the final read-out electronics for ATLAS which measures both the drift time and the pulse height
of ionizing particle tracks. The chamber was operated at photon count rates of up to 990~Hz/cm$^2$.
A silicon strip detector telescope (see Figure~\ref{fig2}) 
was used as external reference to determine the space-to-drift time
relationship and the spatial resolution and efficiency of the drift tubes.


\begin{figure}[t!]
\begin{center}
\includegraphics[width=\linewidth]{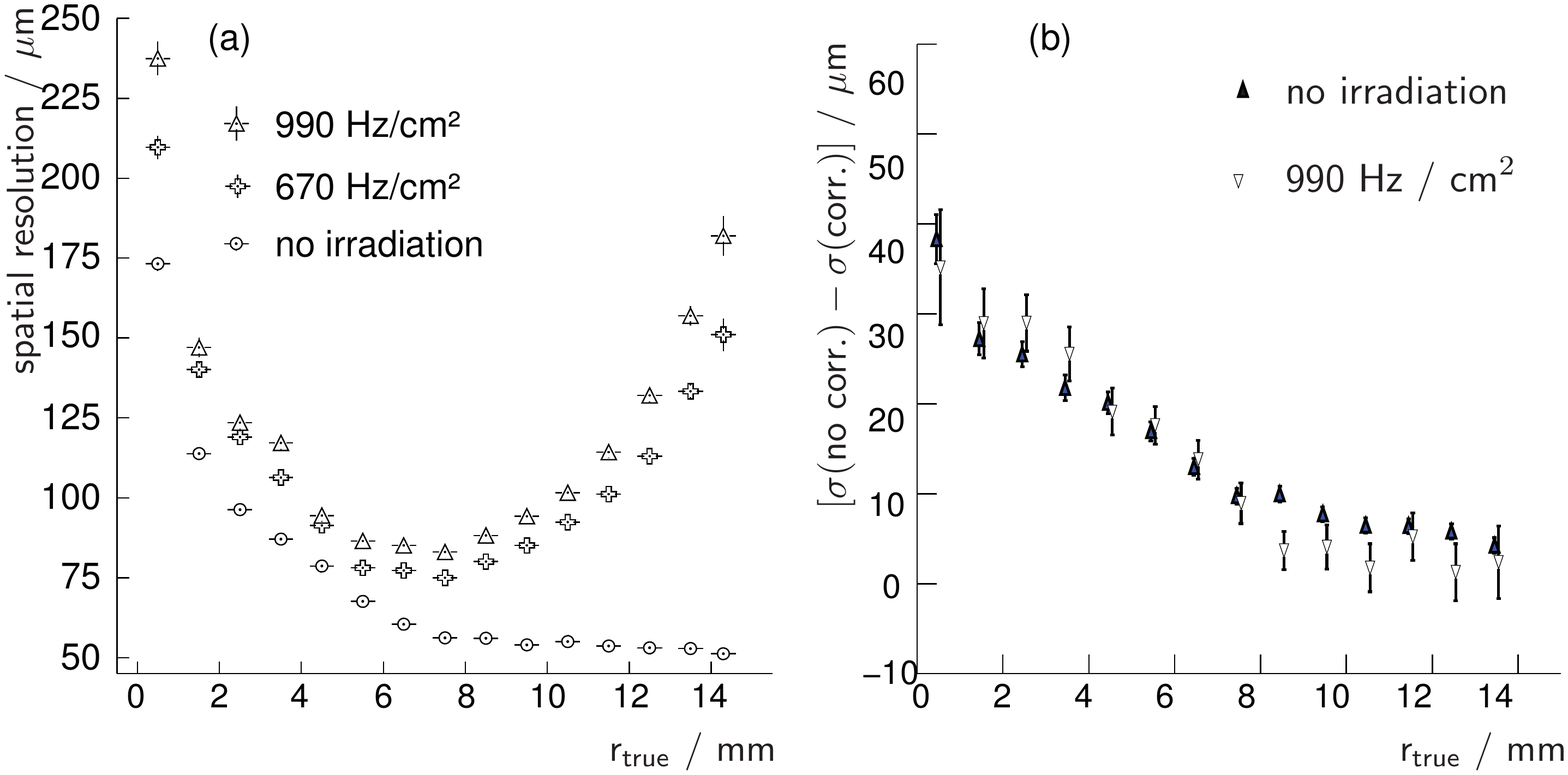}
\caption{\label{fig3}a) The spatial resolution of the drift tubes as a function of the
impact radius $r_{\mathrm true}$ of the muon determined by the silicon detector telescope
at different photon count rates after time-slewing corrections.
b) Improvement of the average drift-tube resolution due to time-slewing corrections
using the pulse-height measurement of the read-out electronics. The improvement increases
at smaller impact radii and is independent of the irradiation rate.}
\end{center}
\end{figure}

\begin{figure}[t!]
\begin{center}
\includegraphics[width=\linewidth]{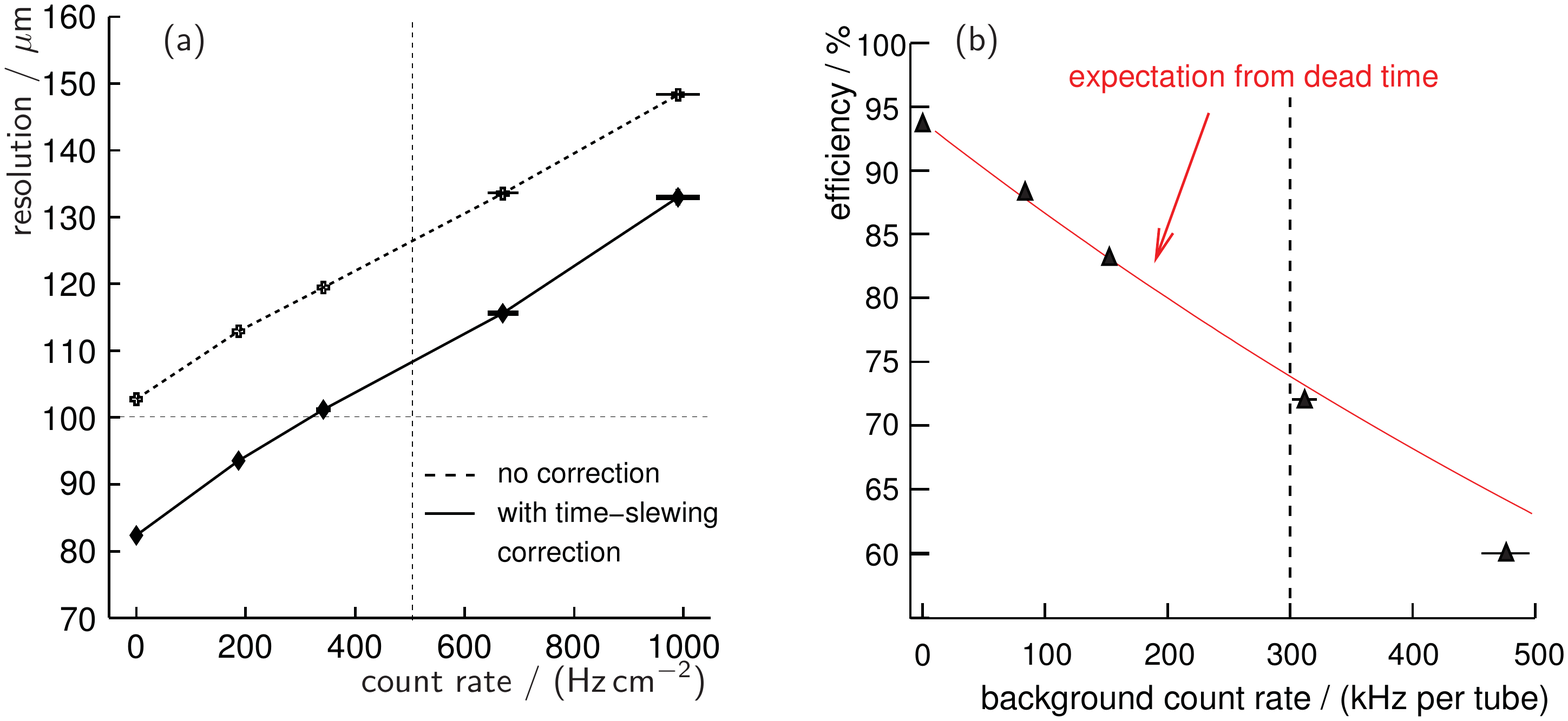}
\caption{\label{fig4}a) The average spatial resolution of the drift tubes with and without
time-slewing corrections and b) the drift-tube efficiency
for 790~ns dead time as a function of the photon count rate.}
\end{center}
\end{figure}


\section{Spatial Resolution Measurement\label{section2}}
The discriminator threshold of the read-out electronics was optimized
to a value of 4.6 times the thermal noise fluctuations
corresponding to the 25$^{\mathrm th}$ primary ionization electron.
By extrapolating the muon trajectory measured by the silicon strip
detector telescope with $10~\mu$m accuracy to the nearest triple layer of the
chamber at a distance of 58~cm, the muon impact radius $r_{\mathrm true}$
in the drift tubes of this layer is determined with a precision of $20~\mu$m.
With this information, the space-to-drift time relationship and the spatial
resolution of the drift tubes as a function of the impact radius is determined
depending on the $\gamma$ irradition rate (see Figure~\ref{fig3}a).

The measured drift radii show a systematic deviation from the impact radii
determined with the silicon detector telescope which increases with decreasing
pulse height measured by the read-out electronics. The deviation is interpreted
as a time-slewing effect of the read-out electronics. By parametrizing the
observed correlation as a function of pulse-height and impact radius, an
average time-slewing correction function is determined. The time-slewing
corrections lead to a significant improvement of the drift-tube resolution
of up to $40~\mu$m at small radii independent of the irradiation rate
(see Figure~\ref{fig3}b).

The Ar:CO$_2$ gas mixture shows a strong dependence of the drift velocity 
on the electric field. Therefore, fluctuations in the space charge density 
created in the tubes by the $\gamma$ irradiation cause an uncertainty in the 
space-to-drift time relationship. This effect~\cite{aleksa1},\cite{MPI2}
leads to a degradation of the spatial resolution with increasing 
irradiation rates (see Figure~\ref{fig3}a). The degradation increases rapidly
for large impact radii. The average drift tube resolution as a function of the
photon count rate is shown in Figure~\ref{fig4}a with and without time-slewing 
corrections. The resolution degrades linearly with increasing count rate.

Without irradiation, the average drift-tube resolution
is $104~\mu$m without and $82~\mu$m with time-slewing corrections.
For the maximum expected background count rate in ATLAS, 500~Hz/cm$^2$, the
time-slewing corrections improve the resolution from $128~\mu$m to $108~\mu$m.

\section{Drift-Tube Efficiency\label{section3}}
Because of the fixed dead time (790~ns in the test beam measurements) 
built into the read-out electronics, muon hits
in the drift tubes can be masked by earlier hits of $\delta$-rays and from the
background radiation. Hence, a reduction of the drift tube efficiency by $6\%$
due to $\delta$-rays is expected even without irradiation and
a further decrease of efficiency with increasing irradiation rate.

The probability of finding a hit with drift radius compatible with
the muon impact radius determined by the silicon detector telescope
within three times the spatial resolution is shown in Figure~\ref{fig4}b. 
Up to the highest anticipated
count rate per tube in ATLAS of 300~kHz, the measured efficiency
follows the expectation based on a simple model of the read-out electronics
with 790~ns dead time without taking into account details of the pulse shape.
The muon hit efficiency drops from $94\%$ without photon irradiation to $72\%$
at 300~kHz background count rate per tube.
Due to the redundant track-point measurements in the 6 to 8 tube layers of a 
MDT chamber, the measured drift-tube efficiency allows for an efficient 
reconstruction of muon trajectories in the ATLAS muon spectrometer up to the
highest expected background rates.

\section{Conclusions\label{section4}}
A full-scale monitored drift-tube chamber from the serial production for the
ATLAS muon spectrometer equipped with final read-out electronics has been tested 
in a muon beam at $\gamma$ irradiation rates of up to twice the maximum background 
rate expected during operation at the LHC. 
Even at the highest anticipated background rate of 500~Hz/cm$^2$ which leads to a
deterioration of the drift tube resolution by $25~\mu$m, 
a spatial resolution of close to $100~\mu$m is achieved as required 
by applying time-slewing corrections. The measured drift tube efficiency as a function
of the photon count rate follows the expectation from the built-in dead time of the
read-out electronics.


\end{document}